\documentclass[pre,longeq]{revtex4}

\usepackage{color}
\usepackage{graphicx}
\usepackage{color}
\graphicspath{{./figures/}}

\begin{document}
\title{Signature of jamming under steady shear in dense particulate suspensions}
\author{Subhransu Dhar}
\author{Sebanti Chattopadhyay}
\author{Sayantan majumdar}
\email[]{smajumdar@rri.res.in}
\affiliation{Raman Research Institute, Bangalore 560080, India}
\date{\today}
\begin{abstract}
Under an increasing applied shear stress ($\sigma$), viscosity of many dense particulate suspensions increases drastically beyond a stress onset ($\sigma_0$), a phenomenon known as discontinuous shear-thickening (DST). Recent studies point out that some suspensions can transform into a stress induced solid-like shear jammed (SJ) state at high particle volume fraction ($\phi$). SJ state develops a finite yield stress and hence is distinct from a shear-thickened state. Here, we study the steady state shear-thickening behaviour of dense suspensions formed by dispersing colloidal polystyrene particles (PS) in polyethylene glycol (PEG). We find that for small $\sigma$ values the viscosity of the suspensions as a function of $\phi$ can be well described by Krieger-Dougherty (KD) relation. However, for higher values of $\sigma$ ($>> \sigma_0$), KD relation systematically overestimates the measured viscosity, particularly for higher $\phi$ values. This systematic deviation can be rationalized by the weakening of the sample due to flow induced failures of the solid-like SJ state. Using Wyart-Cates model, we propose a method to predict the SJ onset from the steady state rheology measurements. Our results are further supported by in-situ optical imaging of the sample boundary under shear. 
\end{abstract}

\maketitle

\section{Introduction}
Dense suspensions formed by dispersing hard particles in a liquid at high volume fractions ($\phi$) (typically, $> 0.5$) show an array of interesting non-linear stress response \cite{Brown2009, Fall2015, Hermes2016}. At sufficiently large $\phi$ values, many dense suspensions show discontinuous shear-thickening (DST) \cite{Barnes1989, Hoffman1972, Bender1996, Maranzano2001, Fall2008, Xu2014} when the suspension viscosity rapidly increases by more than an order of magnitude under an applied stress ($\sigma$) larger than a stress onset ($\sigma_0$). Upon removal of the applied stress, the viscosity quickly drops and approaches the initial value. This striking effect has triggered a lot of recent interests to study dense suspensions \cite{Guy2015, Royer2015, Comtet2017, Wagner2009, Singh2018}. Such control of suspension viscosity by an external stress has made these systems potential candidates for designing smart and stimuli responsive materials that can find wide range of future applications \cite{Lee2003, Majumdar2013}. 

While, many recent studies find that shear-thickening in dense particulate suspensions are reversible \cite{Maranzano2001, Lin2019, Qin2017}, few others report significant hysteresis effects \cite{Chu2014, Majumdar2011}. For our system (PS particles in PEG), we find that hysteresis effects are negligible for both parallel-plate as well as cone-plate geometries as shown in the supplementary figure (Fig. S1). Such reversibility and instantaneous shear-thickening response is also reported earlier for a similar system \cite{Qin2017}.
   
Hydrodynamic lubrication based models alone are not sufficient to understand strong shear-thickening in dense suspensions \cite{Brown2014}, as the inter-particle friction also plays a key role \cite{Fernandez2013, Lin2015, Clavaud2017}. Recent numerical studies \cite{Seto2013, Mari2015} demonstrate that DST originates from stress induced rapid build-up of frictional contacts when lubrication layers between the particles break down due to enhanced inter-particle pressure at high $\sigma$ values. In recent years, experiments reveal that at very large values of $\phi$, approaching the random close packing ($\phi_{rcp}$), many of these systems get into a shear induced solid-like state known as shear jammed (SJ) state  having a finite yield stress \cite{Majumdar2011, Smith2010, Peters2016, Majumdar2017, James2018, Han2018}. The existence of such solid-like SJ state has been predicted earlier by a phenomenological model by Wyart and Cates (WC) \cite{Wyart2014} based on stress induced enhancement of number of frictional contacts in the system as a function of increasing applied stress beyond a threshold. WC model predicts that for frictional particles, the viscosity of the system will diverge at a stress dependent jamming packing fraction $\phi_{J}(\sigma)$ given by,
\begin{equation}
\label{eq1}
\phi_{J}(\sigma) = f(\sigma)\,\phi_m + [1 - f(\sigma)]\phi_0
\end{equation}
where, $f (\sigma)$ is the fraction of particles making frictional contacts with friction coefficient $\mu > 0$, $\phi_0$ and $\phi_m$ represent jamming packing fractions for $f = 0$ (when all the contacts are lubricated) and $f = 1$ (when all the contacts are frictional), respectively. Dense suspensions are extremely complex due to the presence of large number of degrees of freedom and microscopic interactions. WC model provides microscopic insights of shear-thickening and jamming behaviour in these systems, based on a single stress dependent order parameter $f(\sigma)$. Remarkably, wide range of dense suspensions of frictional particles agrees with WC model despite having very different microscopic details \cite{Guy2015}.    

While steady state viscosity measurements successfully describe the shear-thickening behaviour, they cannot capture  signatures of SJ state \cite{Peters2016}. The main difficulty to probe shear-induced jamming by steady state rheology measurement originates from the fact that in a rigid solid-like SJ state, it is not possible to produce steady-velocity gradients. In this state, the sample can only flow by producing various failures. Such failures give rise to a finite apparent viscosity of a shear jammed state when the real viscosity should tend to infinity. The experimental studies that distinguish the shear-jammed state from a shear-thickened state \cite{Peters2016, James2018, Han2018} measure transient stress response of the suspension. However, studies in references \cite{Peters2016, James2018} also point out that the steady state flow curves do not directly show any signature of shear jamming transition. To our knowledge, there are no studies that experimentally distinguish a SJ state from a shear-thickened state based only on steady state flow curves. 

In this letter, we study shear-thickening behaviour of suspensions formed by colloidal polystyrene (PS) particles over a wide range of $\phi$ and $\sigma$ values. From the steady state flow-curves we obtain systematic deviations of viscosity from the KD relation \cite{Krieger1957} for large $\sigma$ and $\phi$ values. Using such deviations together with WC model (Eq. 1), we predict the onset stress for shear jamming ($\sigma_{SJ}$) for a given packing fraction. Our conclusion of weakening of the sample at high stress and volume fractions (as concluded from the deviations from KD relation) is consistent with the expected failures caused due to the flow of solid-like SJ state. Furthermore, the agreement of our data with WC model brings out the importance of frictional contacts to observe shear-jamming.  Although, both KD relation and WC model have been widely studied in the context of dense suspensions, to our knowledge, they have not been used to determine $\sigma_{SJ}$. Thus, our experiments propose a novel method to distinguish a SJ state from a shear-thickened state, entirely from the steady state measurements.

\section{Experimental}
Suspensions are formed by dispersing PS particles in polyethylene glycol (PEG-400) \cite{Qin2017} at different volume fractions $\phi$ ranging from 0.4 - 0.6. Details of the particle synthesis and SEM imaging are mentioned in the Supplementary Information (S.I). Just before loading the sample for rheology measurements, the suspensions are ultrasonicated for 5 minutes to ensure uniform dispersion. Rheology measurements are carried out using a stress controlled Rheometer (MCR-702, Anton Paar, Austria) with parallel plate (plate diameter of 25 mm and a gap of 300 $\mu$m) and cone and plate (cone diameter of 25 mm and cone angle $\approx$ 2$^{o}$) geometries at a temperature of 25$^{o}$C.

To obtain steady state flow-curves (shown in Fig. 1e), we vary the applied shear stress on the sample with a waiting time per data point varying logarithmically from 20 s (at the lowest stress) to 1 s (at the highest stress). We choose such protocol, because, for lower applied stress values, the shear rate produced in the sample is very small. To measure the viscosity reliably, we need to wait for a sufficient time for the sample to undergo appreciable strain deformation.  On the other hand, higher values of applied stress (deep inside the shear-thickening regime) produces high shear rate. So, the sample already undergoes significant strain deformation within a very short time. Under high shear rate, we find that there is a tendency of the sample to gradually protrude out of the rheometer plates that can result in measurement artefacts. We also vary the waiting time per data point over a range as shown in the S.I. Fig. S1. We obtain almost the same flow-curves for all the waiting times, signifying that waiting time per data point for our experiment is sufficient for the system to reach a steady state.
\begin{figure*} 
\begin{center}
\includegraphics[width = 13cm]{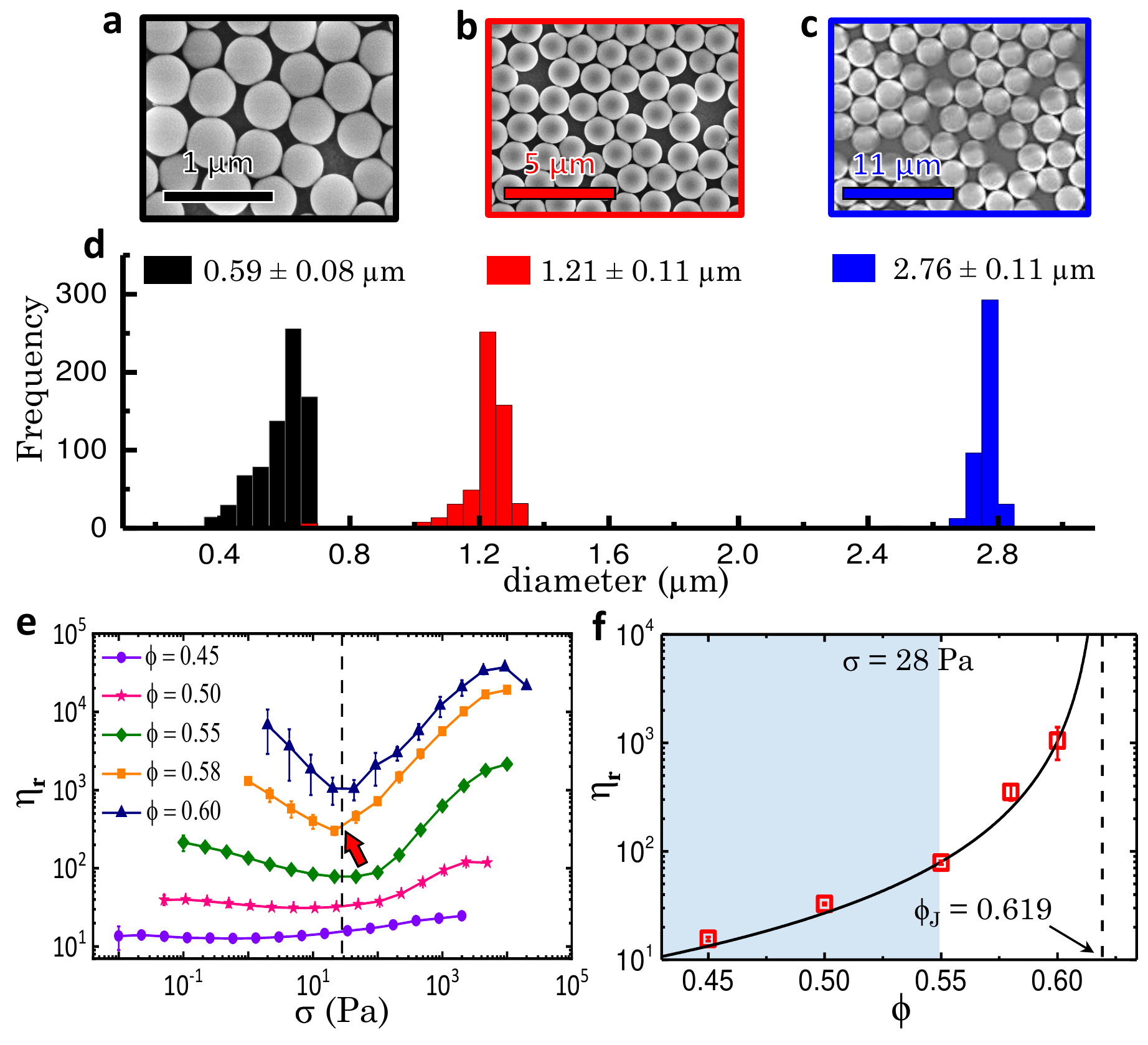}
\caption{\label{F1} (a) - (c), SEM images of synthesized PS microspheres for three different sizes. (d) Histograms indicating the distribution of particle size with mean diameter ($d$) and standard deviation indicated in the figure. (e) Relative viscosity ($\eta_r$) as a function of applied shear stress ($\sigma$) for particles with $d$ = 1.21 $\mu$m for different volume fractions $\phi$. The onset stress ($\sigma_0$) for $\phi$ = 0.58 is marked with a bold arrow. The error bars are the standard deviations of viscosity for two independent measurements. (f) $\eta_r$ vs $\phi$ for the value of $\sigma$ indicated by the dashed line in panel (e). The solid-line in (f) indicates a fit of KD relation to the three data points for $\phi$ = 0.45, 0.5 and 0.55 (shaded region). The vertical dashed-line indicates the value of $\phi_J$ (described in the main text). 
}
\end{center}
\end{figure*}
\begin{figure*} 
\begin{center}
\includegraphics[width = 13cm]{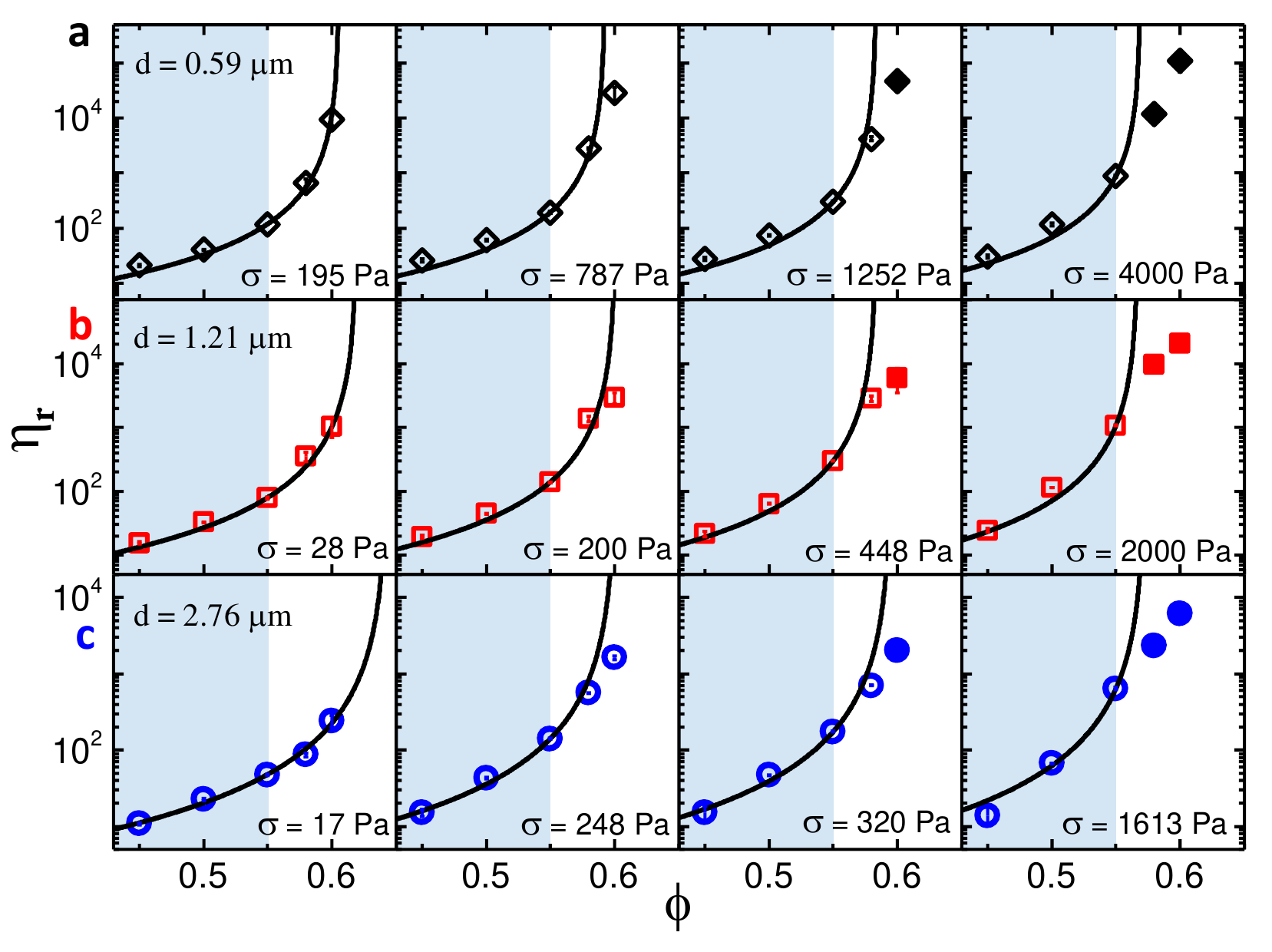}
\caption{\label{F2} Relative viscosity ($\eta_r$) as a function of volume fraction for different applied stress ($\sigma$) values as indicated. Each row represents a particular mean size of the particles; (a) $d$ = 0.59 $\mu$m, (b) $d$ = 1.21 $\mu$m, (c) $d$ = 2.76 $\mu$m. The solid line is a fit to KD relation based on the values of $\phi$ in the shaded region. For all three sizes of the particles, we obtain significant deviation (indicated by solid symbols) from KD fits for large $\sigma$ and $\phi$ values. The error bars are the standard deviations of viscosity for two independent measurements and are small compared to the size of the symbols in most cases.
}
\end{center}
\end{figure*} 
\section{Results and discussions}
Typical Scanning Electron Microscopy (SEM) images of the synthesized particles (Fig. 1a - c) along with the size distributions (Fig. 1d) are shown. Flow curves, showing the relative viscosity ($\eta_r = \frac{\eta_{suspension}}{\eta_{solvent}}$) as a function of applied shear-stress ($\sigma$) at different packing fractions ($\phi$) are plotted in Fig. 1e for almost six orders of magnitude of stress variation for mean particle diameter $d$ = 1.21 $\mu$m. After an initial Newtonian/shear-thinning region, there is a shear-thickening region at higher values of $\sigma$ beyond a stress onset $\sigma_0$, when $\eta_r$ increases strongly for $\phi \geq$0.55. The onset stress ($\sigma_0$) is given by the minimum applied stress for which $\frac{d~log(\eta_r)}{d~log(\sigma)} >$ 0 (indicated in Fig.1e with a bold arrow). Physically, $\sigma_0$ indicates the minimum stress scale for inducing frictional contacts between two particles in the suspension, overcoming the residual inter-particle interactions \cite{Brown2014}. We see from Fig. 1f that $\eta_r$ increases monotonically with increasing $\phi$ values at a particular value of $\sigma$ ( = 28 Pa, as indicated by the vertical dashed-line in Fig. 1e). Such non-linear increase in viscosity as a function of volume fraction can be described by the empirical Krieger-Dougherty (KD) model \cite{Krieger1957, Wyart2014} for spherical particles: $\eta_r = (1 - \frac{\phi}{\phi_J})^{-2}$. This model predicts that as the particle volume fraction ($\phi$) gradually increases, the viscosity ($\eta_r$) also increases and finally diverges when $\phi$ closely approaches $\phi_J$, the jamming packing fraction. To avoid the complications due to failures originating from the flow of solid-like SJ state, we restrict our fitting up to the random loose packing $\phi_{rlp}$ ($\sim$ 0.55) of the system, below which shear induced jamming is not expected \cite{Silbert2010}. So, we fit the data only over the range of packing fractions $0.45 \leq \phi \leq 0.55$ (shaded region) in Fig. 1f. We also compare the $\phi_J$ values obtained from this fitting range to that obtained from a slightly larger range $0.4 \leq \phi \leq 0.55$ for different $\sigma$ values (S.I. Fig. S2). We get identical values of $\phi_J$ in all cases. Importantly, for smaller values of $\phi$, the shear rate produced in the sample becomes very high at higher values of applied stress and the sample tends to come out of the rheometer plates. For this reason, we cannot extend the range down to $\phi$ = 0.4 for very high stress values (S.I. Fig. S2).
For the rest of the manuscript, we will stick to the range $0.45 \leq \phi \leq 0.55$ for fitting KD model. $\eta_r$ vs $\phi$ shows a very good agreement with KD model as shown by the fitted solid-line in Fig. 1f, with $\phi_J = 0.619$.

Next, using the flow curve data ($\eta_r$ vs $\sigma$) for different $\phi$ values, we show the variation of $\eta_r$ as a function of $\phi$ for different $\sigma$ values in Fig. 2. For simplicity, we have shown such flow curve data only for $d$ = 1.21 $\mu$m in Fig. 1e. In S.I. Fig. S3, we explain the detailed procedure to obtain $\eta_r$ vs $\phi$ for different $\sigma$ values from the flow-curves. In Fig. 2a, we show $\eta_r$ vs $\phi$ for $d$ = 0.59 $\mu$m when $\sigma$ is gradually increased from the left most to the right most panel. For a given value of $\sigma$ we fit KD model to the data over the shaded region ($0.45 \leq \phi \leq 0.55$). We obtain very good agreement for lower $\sigma$ values, however, with increasing $\sigma$, the fits systematically deviate from the experimentally measured data points for higher values of $\phi$. Similar trends are also observed for $d$ = 1.21 $\mu$m (Fig. 2b) and $d$ = 2.76 $\mu$m (Fig. 2c) particles, albeit, over different stress ranges. This indicates that for large $\sigma$ and $\phi$ values, the experimentally measured $\eta_r$ is much lower compared to the KD predictions. We interpret that these deviations indicate shear-induced solidification of the suspension where steady velocity gradient cannot be maintained without failures in the sample. Here, we use the term `failure' in a very general sense. In disordered materials brittle-fracture, shear band plasticity, localized non-affine deformations are the most likely mechanism for yielding/failure \cite{Nicolas2018}. However, with our present imaging set-up (resolution of $\sim$ 50 pixels/mm), we can only observe extreme failures like a crack formation due to brittle failure at very high applied stress values (see Fig. 3e, 3f and S.I. Movie) and not the more subtle ones. Such failures weaken the stress response of the sample that translates into a lower value of measured viscosity.

We now study stress dependence of jamming packing fraction $\phi_J (\sigma)$ obtained from the fitted KD curves shown in Fig. 2. We find that, with increasing values of $\sigma$, $\phi_J (\sigma)$ decreases as shown in Fig. 3a ($d$ = 0.59 $\mu$m), 3b ($d$ = 1.21 $\mu$m) and 3c ($d$ = 2.76 $\mu$m). Such stress dependence of $\phi_J$ can be well described by WC model (Eq. 1). 
The exact form for $f(\sigma)$ can not be determined experimentally. Motivated by earlier studies \cite{Howell1999, O'Harn2001}, an exponential form for $f(\sigma)$ is introduced \cite{Guy2015}. However, it is found that a stretched exponential form, $f(\sigma) = e^{-(\sigma^{*} / \sigma)^{\beta}}$ gives better agreement with the experimental data \cite{Guy2015}. The exponent $\beta$ is a fitting parameter that gives the range of stress values over which shear-thickening is observed. Higher the value of $\beta$, lower will be the stress range for shear-thickening.
The solid lines in Figs. 3a, 3b and 3c indicate the fits of WC Model to the experimental data, where we got an excellent agreement. This observation reconfirms the microscopic picture of stress induced conversion of the lubricated contacts into the frictional ones. For $d$ = 0.59 $\mu$m, 1.21 $\mu$m and 2.76 $\mu$m we got the respective $\beta$ values of 1.19, 0.99 and 0.74. Such trend indicates that the range of stress values for shear-thickening decreases with decreasing particle size. For a typical value of $\beta \sim 1$, an applied stress $\sigma = \sigma^{*}$ gives $f(\sigma) = 1/e$. Thus, $\sigma^{*}$ gives a stress scale when the system develops sufficient number of frictional contacts and is proportional (but not equal) to the average onset stress ($\sigma_0$) for shear-thickening \cite{Guy2019}. Like the onset stress, $\sigma^{*}$ also shows similar dependence on particle size \cite{Guy2015}: $\sigma^{*}\sim 1/d^2$ (S.I. Fig. S4).
\begin{figure*}
\begin{center}
\includegraphics[width = 13cm]{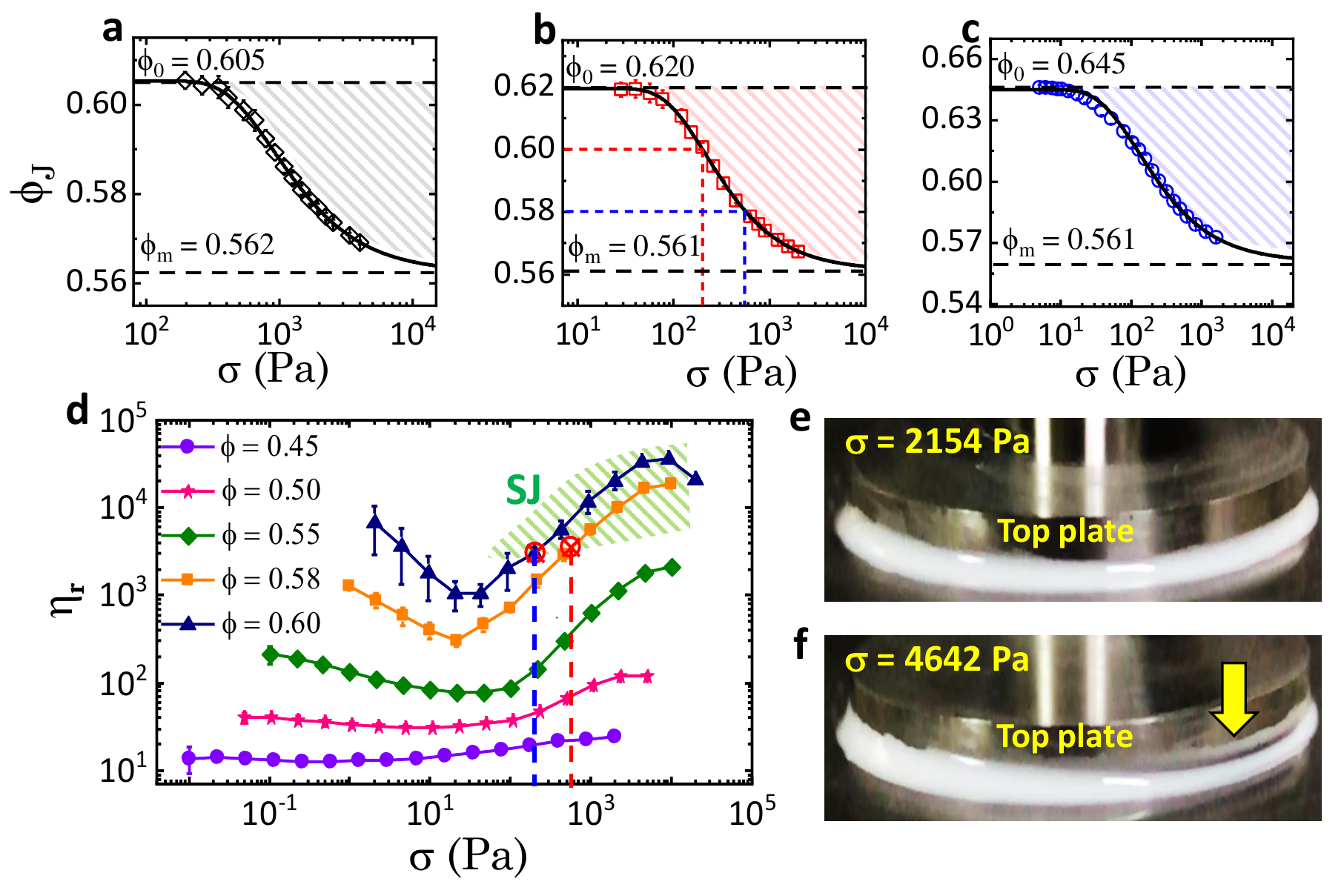}
\caption{\label{F3} Jamming packing fraction ($\phi_J$) as a function of applied stress ($\sigma$) for different particle sizes: (a) $d$ = 0.59 $\mu$m, (b) $d$ = 1.21 $\mu$m, (c) $d$ = 2.76 $\mu$m. The solid lines are the fit to WC model. Shaded region predicts the parameter range for SJ (mentioned in the main text). (d) Steady state flow curves for $d$ = 1.21 $\mu$m. The shaded region indicates the SJ regime with $\sigma_{SJ}$ is indicated for $\phi$ = 0.58 (blue dashed-line) and $\phi$ = 0.6 (red dashed-line), obtained from (b). (e) and (f) show optical images of the sample boundary between the rheometer plates inside SJ regime, for $\phi$ = 0.58. For very high stress values, partial detachment of the sample from the plate (due to brittle failure) is observed (as indicated by the bold arrow). The aspect ratio is changed from the original images for clarity. 
}
\end{center}
\end{figure*}
Physically, WC model predicts the packing fraction of particles $\phi_{J}(\sigma)$ for a given applied stress $\sigma$ at which the viscosity of the suspension $\eta_r (\sigma)$ diverges. It implies that, for a given $\sigma$ value, the suspension having packing fraction $\phi$ ($> \phi_{J}(\sigma)$) should behave like a jammed solid with infinite viscosity. Thus, existence of finite viscosity for the parameter range marked by the shaded regions in Figs. 3a, 3b and 3c is unphysical. We conjecture that such finite value of viscosity at high $\sigma$ and $\phi$ values can be observed (e.g. Fig. 1e, Fig. 2) if the sample develops flow induced failures. For a given $\phi$, the onset stress for shear jamming ($\sigma_{SJ}$) is determined from the stress value $\sigma$ for which, $\phi = \phi_{J}(\sigma)$ (shown in Fig. 3b by dashed vertical lines for $\phi$ = 0.58 and 0.6). We show the SJ region (shaded region) for $d$ = 1.21 $\mu$m in Fig. 3d with $\sigma_{SJ}$ values are indicated for $\phi$ = 0.58 and 0.6. We find that $\sigma_{SJ}$ decreases for increasing $\phi$. Such trend has also been reported for SJ under transient forcing \cite{Peters2016, Han2018}. We perform in-situ optical imaging of the sample in the flow-gradient (\textbf{v}, $\nabla$v) plane (S.I.Movie) while measuring the flow-curves. Since, the particles are optically opaque, we can only image the sample boundary. Deep inside the SJ regime, we indeed find the detachment and edge fracture of the sample (Fig. 3e, 3f and S.I.Movie), reconfirming the formation of solid-like SJ state under shear.

For $d$ = 2.76 $\mu$m (Fig. 3c), we find that the value of $\phi_0$ (indicated in Eq. 1) approaches the random close packing ($\phi_{rcp}\approx$ 0.64) for a system of monodisperse hard spheres \cite{Wyart2014, Silbert2010}. We also observe that $\phi_0$ systematically drops from $\phi_{rcp}$ value as the particle size decreases (Fig. 3a, 3b). This indicates a deviation from  ideal hard sphere behaviour due to residual inter-particle interactions \cite{AntonPaar}. Although, we do not  know the exact nature of such interactions, the systematic drop from $\phi_{rcp}$ value with decreasing particle size indicates that such interactions are predominantly induced by particle-surfaces \cite{Maranzano2001, Maranzano2001_1}, since, for a given packing fraction, the smaller particles will have a larger surface to volume ratio. This observation is also consistent with the decrease in shear-thickening range for smaller particle sizes (higher values of $\beta$) that indicates enhanced inter-particle interactions \cite{Brown2010}. The parameter $\phi_m$ corresponds to the minimum volume fraction required for shear induced jamming. We find that in our study $\phi_m\approx$ 0.56 in all cases. This value of $\phi_m$ is very close to random loose packing ($\phi_{rlp} \sim$ 0.55) for monodisperse hard spheres \cite{Silbert2010}. This observation also agrees with the reported lower-limit for observing jamming in frictional systems \cite{Silbert2010}. Also, the similar values of $\phi_m$ for all the particle sizes indicate that despite the differences in the residual inter-particle interactions, the stress-induced frictional interactions between the particles are very similar.
\section{Conclusions}
In conclusion, we propose a novel method to predict the onset of shear jamming in dense particulate suspensions entirely based on steady state rheological measurements. We generalize our results for a range of particle sizes. In all cases, the predicted onset stress values for shear jamming ($\sigma_{SJ}$) are found to decrease with increasing $\phi$, a trend also obtained from the transient measurements \cite{Peters2016, Han2018}. In our case, the optical imaging can not detect any signature of sample failure for stress values just beyond $\sigma_{SJ}$ (predicted from our analysis) as shown in the S.I.Movie. We believe that our analysis captures the weakening of the sample due to microscopic failures in random locations of the sample. However, these failures are only visible when they merge and grow to macroscopic length-scales under increasing stress. This implies that our analysis is more sensitive in picking up the onset of SJ state as compared to our optical imaging method probing failures in the sample. Since, the resolution of our imaging is not very good, we can only see extreme failures (like brittle fracture, macroscopic crack formation etc.) but not plastic shear bands and other subtler features. Further, due to the opaque nature of the sample, the failures happening inside the bulk of the sample (not on the surface) will not be optically visible. The generality of our analysis for systems with different particle shapes and interactions needs to be verified. Furthermore, comparison of the predicted value of $\sigma_{SJ}$ from our steady state measurements with that obtained from transient measurements or directly by more sophisticated optical/non-optical imaging techniques probing sample failures, remain an interesting future challenge. We hope that our experiments will motivate further studies on shear induced jamming in dense suspensions.   
\section{Acknowledgments}
SM thanks SERB (under DST, Govt. of India) for a Ramanujan Fellowship. We thank K.M. Yatheendran for help with SEM imaging and Sachidananda Barik, Swaminathan K., Sanjay Kumar Behera for help with particle synthesis.  
\newline

\newpage
{\LARGE{\textbf{Supplementary Information: Signature of jamming under steady shear in dense particulate suspensions}}}
\vspace{4pt}

\vspace{2 cm}
\large{\textbf{Movie description}} 
\vspace{0.5 cm}  

In-situ deformation of the sample boundary in the flow-gradient (\textbf{v}, $\nabla$v) plane is captured during the steady state flow-curve measurement (Fig. 3d, main text) as shown in S.I.Movie. The movie is captured by a digital camera (iPhone X) with a resolution of 1920 X 1080 pixels at frame rate of 60 Hz with a spatial resolution of $\sim$ 50 pixels/mm. The optically opaque nature of the sample enables us to track the boundary deformations as a function of increasing applied stress values. At large values of applied stress (indicated in the movie), there is a clear signature of detachment of the sample from the shearing plate. However, as the applied stress is removed such failure starts to recover when the sample is again transformed back to a liquid like state from a shear jammed state.
\vspace{1 cm}  

\large{\textbf{Synthesis of polystyrene particles}}
\vspace{0.5 cm}

The monodisperse polystyrene (PS) microspheres are synthesized by dispersion polymerization method \cite{Wang2013, Paine1990, Cho2016}. Polyvinylpyrrolidone (PVP K-30, Spectrochem, India) is mixed with a solvent (ethanol/ethanol-water/ethanol with 2-methoxyethanol) inside a 250 ml round bottom flask under continuous stirring at 300 rpm. In another beaker, polystyrene monomers (TCI, Japan), were mixed with the initiator AIBN (Spectrochem, India). The mixture containing the monomers is poured into the flask maintained at 70 $^{o}$C under a constant nitrogen environment. The reaction is carried out for 24 h. The  amount of different reactants controls the particle size \cite{Paine1990, Cho2016}, as summarized in Table \ref{T1}. The PS microspheres formed are cleaned repeatedly using ethanol and water mixture to get rid of chemical impurities/unreacted components from the surface of the particles and are dried and stored for further use.

\begin{table}[h!]
\centering
\caption{\label{T1}Effect of the amount of reactants and solvent type on particle size}
\footnotesize
\begin{tabular}{@{}lllllll}


PS (g) &\ PVP (g) &\ AIBN (g)&\ Ethanol (g)&\ Water (g)&\ 2-methoxyethanol (g)&\ $d$ ($\mu$m)\\
\\
14.7 &\	1.44 &\	0.4 &\ 78.56&\ -- &\	-- &\ 2.76\\ 15.13 &\	1.75 &\	0.24 &\ 27.48&\ --&\	62.73&\ 1.21\\ 8 &\	2 &\ 0.114 &\ 62.82	&\ 19.94	&\ -- &\ 0.59\

\end{tabular}\\
\end{table}
\normalsize
\vspace{1 cm}

\large{\textbf{Sample preparation for SEM imaging}}
\vspace{0.5 cm}

Size distributions of PS microspheres are characterized by scanning electron microscopy (SEM) technique, using  a Ultra Plus FESEM (Zeiss, Germany). First, a small amount of dried PS microspheres is dispersed in water to form a very dilute suspension. The suspension is then ultrasonicated to break particle clusters, if present. A small drop of this suspension is put on top of an ITO coated glass plate using a micro-pipette and is left to dry overnight in a covered petri dish to prevent dust accumulation. The slow drying of the dilute suspension ensures formation of mono-layers of particles over a few localized regions on the ITO plate. Since, PS particles are non-conducting, the ITO plate containing the dried suspension droplet is coated with a thin layer of platinum (2 - 3 nm) to increase the conductivity of the sample, as required for SEM imaging. This platinum coated sample is then mounted in the sample chamber of the SEM and is imaged under vacuum. 

\newpage

\begin{figure*} 
\begin{center}
\includegraphics[width = 15cm]{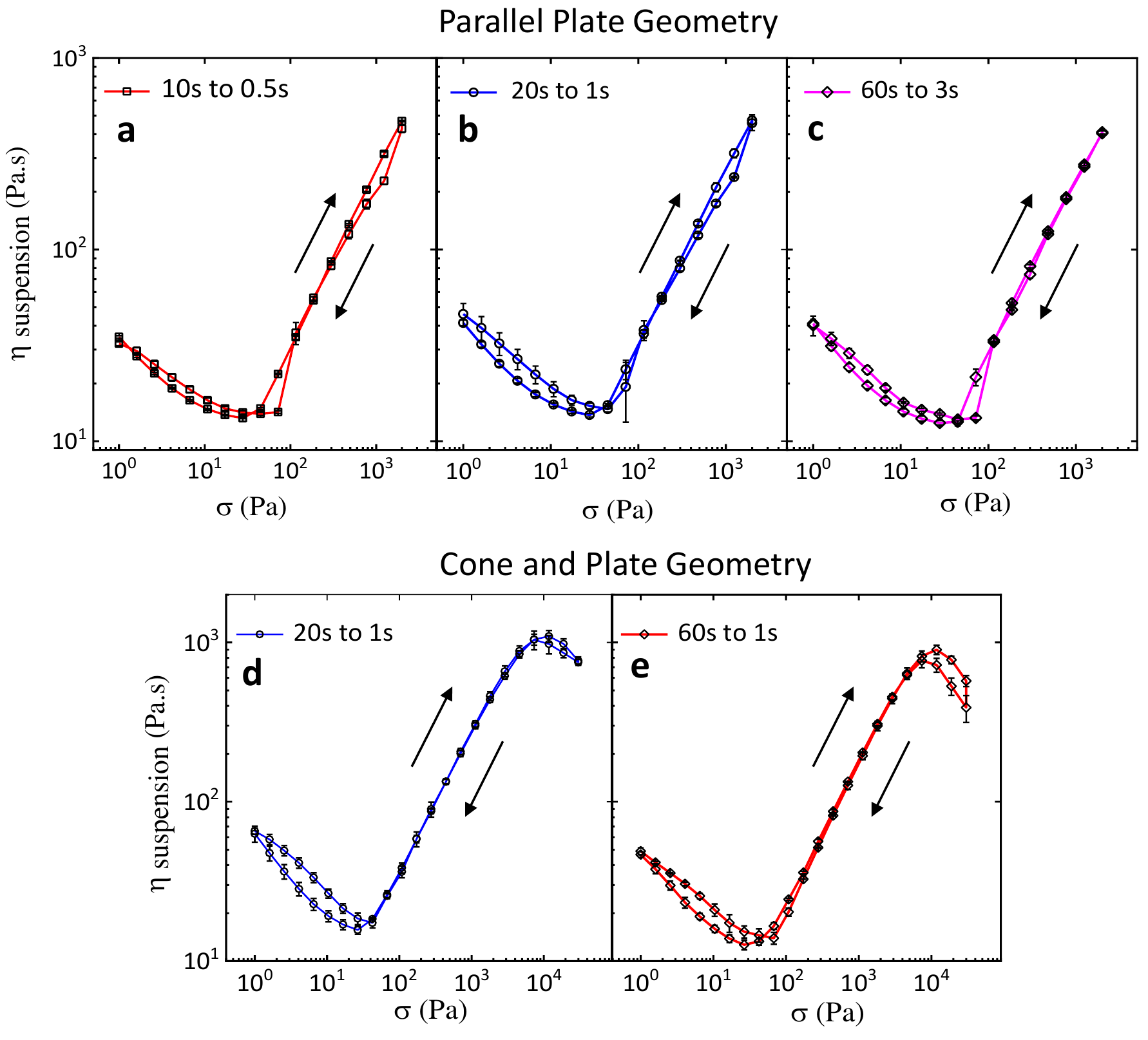}
\renewcommand{\thefigure}{S1}
\caption{\label{FS1} Reversibility and waiting-time dependence of shear-thickening of PS in PEG system ($\phi$ = 58 \%, $d$ = 1.21 $\mu$m). (a), (b) and (c) indicates suspension viscosity ($\eta_{suspension}$) as a function of applied stress ($\sigma$) for both increasing and decreasing stress values for different waiting-time per data point, as indicated (also mentioned in the main text). These measurements are done with a parallel-plate (P-P)  geometry. (e) and (f) indicate the same measurements with a cone and plate (C-P) geometry. We observe that the shear-thickening behaviour is very similar (in both the geometries) for increasing and decreasing stress values, indicating negligible hysteresis effect for this system. Further, the shear-thickening response does not show any waiting time-dependence, signifying a steady state behaviour of the system. In all cases, the $\eta_{suspension}$ vs $\sigma$ curves for increasing stress values are averaged over three experimental runs. Similarly, for decreasing stress values the curves are also averaged over the three runs. The error bars are the corresponding standard deviations.
}
\end{center}
\end{figure*} 

\begin{figure*} 
\begin{center}
\includegraphics[width = 11cm]{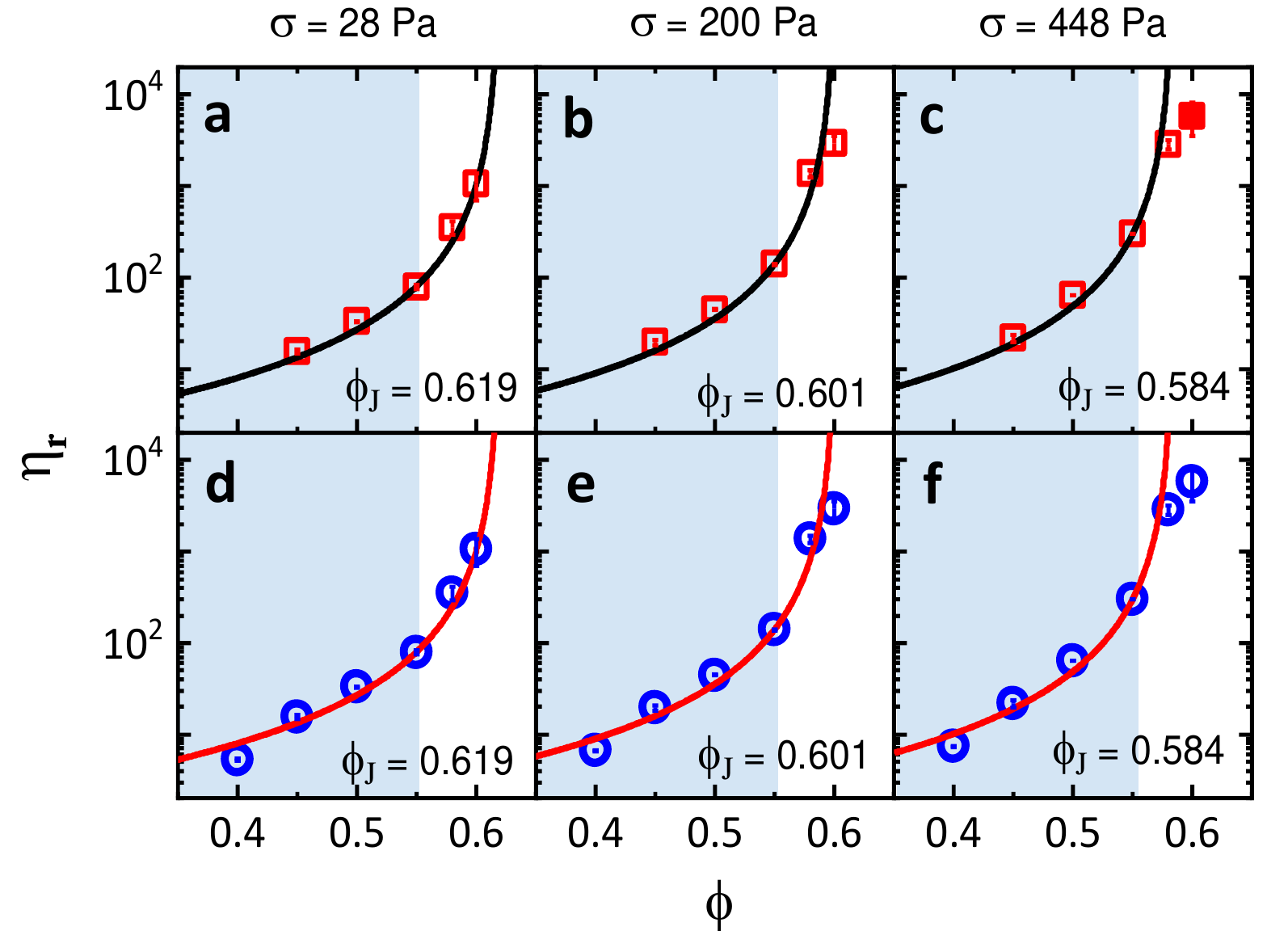}
\renewcommand{\thefigure}{S2}
\caption{\label{FS2} Comparison of the value of jamming packing fraction $\phi_J$ as mentioned in the main text) for different range of fitting. (a), (b) and (c) show the fits to KD equation (described in the main text) for $\sigma$ values obtained over the range $0.45 \leq \phi \leq 0.55$. Same fitting is shown over the range $0.4 \leq \phi \leq 0.55$ in (d), (e) and (f). The fitting parameters ($\phi_J$) obtained for a given $\sigma$ value are same (up to at least 3 decimal places) for both the fitting ranges (as indicated in the figure). Importantly, for $\phi$ = 0.4, the suspension viscosity being relatively low, very high applied stress ($\sigma >$ 1000 Pa) produces high shear rates. Under such high shear rates, the sample tends to protrude out of the rheometer plates. For this reason, we cannot get a reliable viscosity value for $\phi$ = 0.4 for $\sigma >$ 1000 Pa. So, we can not compare the data for $\sigma$ = 2000 Pa (shown in Fig. 2b in the main text for $0.45 \leq \phi \leq 0.55$). 
}
\end{center}
\end{figure*} 

\begin{figure*} 
\begin{center}
\includegraphics[width = 14cm]{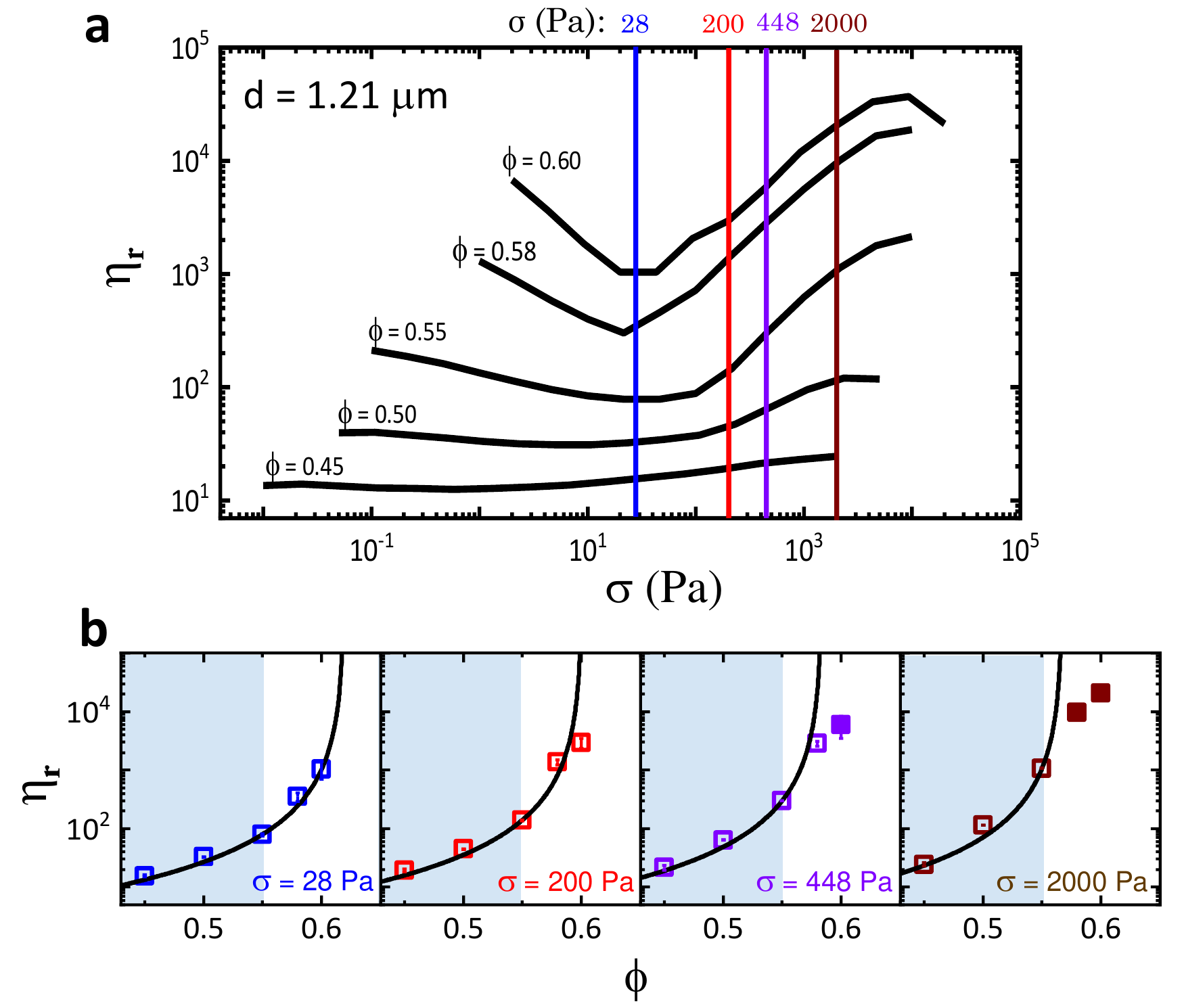}
\renewcommand{\thefigure}{S3}
\caption{\label{FS3} Detailed protocol for obtaining $\eta_r$ vs $\phi$ (indicated in Fig. 2 in the main text) from steady state flow curves. Panel (a) shows steady state flow curves ($\eta_r$ vs $\sigma$) for different $\phi$ values ranging from 0.45 to 0.6 (also shown in Fig. 1e in the main text). We first draw a vertical line for a particular stress value (say, $\sigma$ = 28 Pa), we find the intersection points for the corresponding vertical line (blue) with the flow curves to get $\eta_r$ vs $\phi$, for $\sigma$ = 28 Pa as shown in the left-most panel in (b). We repeat this procedure for different stress values in an increasing order (as indicated with different coloured vertical lines in (a)) to obtain the other panels in (b). In (b), $\eta_r$ vs $\phi$ are shown for increasing $\sigma$  values from left to the right most panels. Here, $d$ = 1.21 $\mu$m (Fig. 2b in the main text). 
}
\end{center}
\end{figure*} 

\begin{figure*} 
\begin{center}
\includegraphics[width = 9cm]{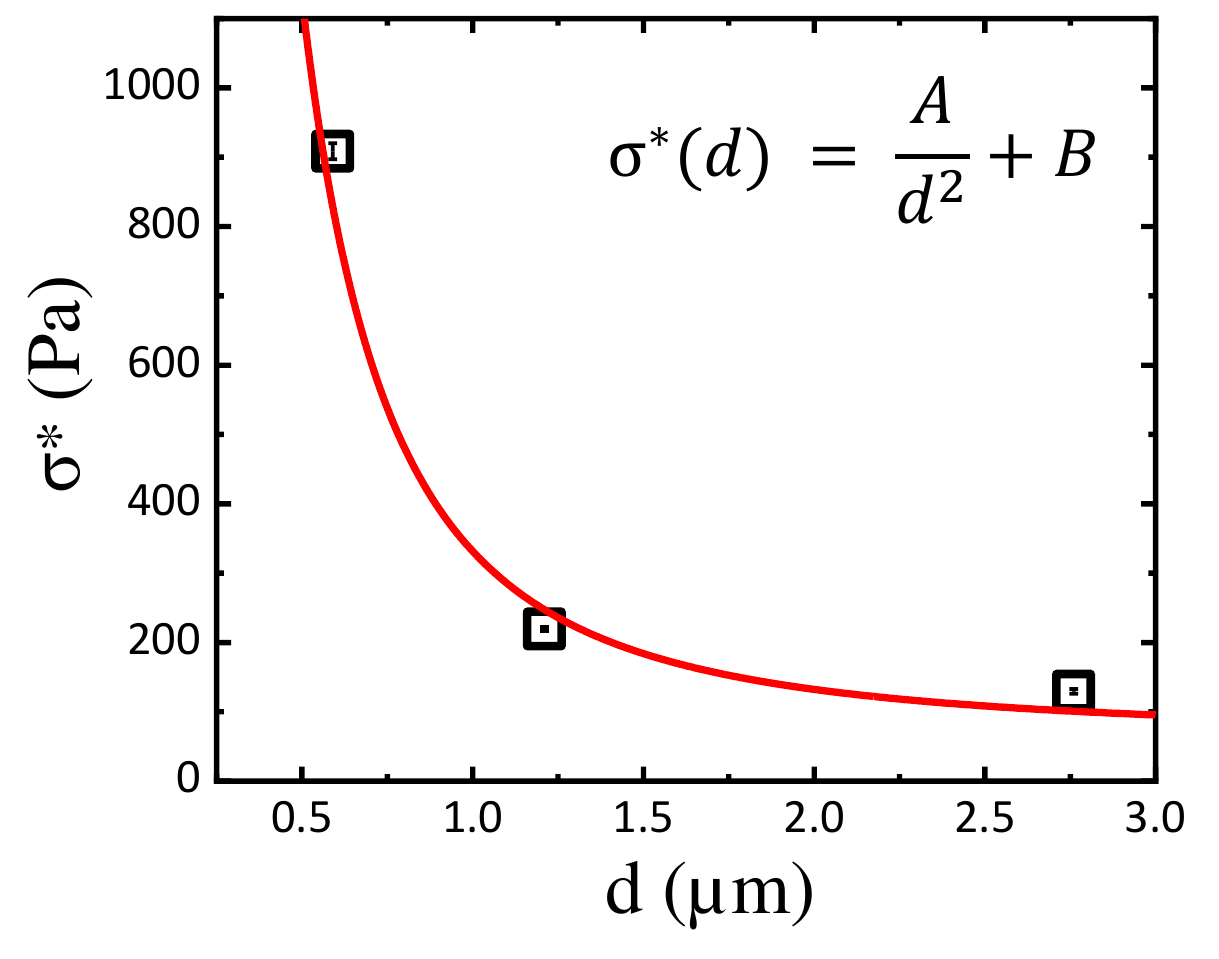}
\renewcommand{\thefigure}{S4}
\caption{\label{FS42} Variation of the parameter $\sigma^{*}$ (mentioned in the main text) with particle size. We get a trend $\sigma^{*} \sim \frac{1}{d^{2}}$. The values of the fitting parameters (indicated in the figure) $A$ = 265.58 $Pa.\mu m^2$ and $B$ = 65.89 $Pa$.
}
\end{center}
\end{figure*}


\begin{thebibliography}{24}
\expandafter\ifx\csname natexlab\endcsname\relax\def\natexlab#1{#1}\fi
\expandafter\ifx\csname bibnamefont\endcsname\relax
  \def\bibnamefont#1{#1}\fi
\expandafter\ifx\csname bibfnamefont\endcsname\relax
  \def\bibfnamefont#1{#1}\fi
\expandafter\ifx\csname citenamefont\endcsname\relax
  \def\citenamefont#1{#1}\fi
\expandafter\ifx\csname url\endcsname\relax
  \def\url#1{\texttt{#1}}\fi
\expandafter\ifx\csname urlprefix\endcsname\relax\def\urlprefix{URL }\fi
\providecommand{\bibinfo}[2]{#2}
\providecommand{\eprint}[2][]{\url{#2}}

\bibitem{Brown2009}
\bibinfo{author}{\bibfnamefont{E.}~\bibnamefont{~Brown}}
  \bibnamefont{and} \bibinfo{author}{\bibfnamefont{H.~M.} \bibnamefont{~Jaeger}},
  \bibinfo{journal}{Phys. Rev. Lett.}
  \textbf{\bibinfo{volume}{103}}, \bibinfo{pages}{086001}
  (\bibinfo{year}{2009}).
	
	\bibitem{Fall2015}
\bibinfo{author}{\bibfnamefont{A.}~\bibnamefont{Fall}},
  \bibinfo{author}{\bibfnamefont{F.}~\bibnamefont{Bertrand}},
	\bibinfo{author}{\bibfnamefont{D.}~\bibnamefont{Hautemayou}},
	\bibinfo{author}{\bibfnamefont{C.}~\bibnamefont{Meziere}},
	\bibinfo{author}{\bibfnamefont{P.}~\bibnamefont{Moucheront}},
	\bibinfo{author}{\bibfnamefont{A.}~\bibnamefont{Lemaitre}},
  \bibnamefont{and} \bibinfo{author}{\bibfnamefont{G.} \bibnamefont{Ovarlez}},
  \bibinfo{journal}{Phys. Rev. Lett.}
  \textbf{\bibinfo{volume}{114}}, \bibinfo{pages}{098301}
  (\bibinfo{year}{2015}).
	
\bibitem{Hermes2016}
\bibinfo{author}{\bibfnamefont{M.}~\bibnamefont{Hermes}},
  \bibinfo{author}{\bibfnamefont{B.~M.}~\bibnamefont{Guy}},
  \bibinfo{author}{\bibfnamefont{W.~C.~K.} \bibnamefont{Poon}},
	\bibinfo{author}{\bibfnamefont{G.}~\bibnamefont{Poy}},
	\bibinfo{author}{\bibfnamefont{M.~E.} \bibnamefont{Cates}},
  \bibnamefont{and} \bibinfo{author}{\bibfnamefont{M.}~\bibnamefont{Wyart}},
  \bibinfo{journal}{J. Rheol.}
  \textbf{\bibinfo{volume}{60}}, \bibinfo{pages}{905}
  (\bibinfo{year}{2016}).

\bibitem{Barnes1989}
\bibinfo{author}{\bibfnamefont{H.~A.}~\bibnamefont{Barnes}},
  \bibinfo{journal}{J. Rheol.}
  \textbf{\bibinfo{volume}{33}}, \bibinfo{pages}{329}
  (\bibinfo{year}{1989}).
	
	\bibitem{Hoffman1972}
\bibinfo{author}{\bibfnamefont{R.}~\bibnamefont{Hoffman}},
  \bibinfo{journal}{Trans. ~Soc. ~Rheol.}
  \textbf{\bibinfo{volume}{166}}, \bibinfo{pages}{155}
  (\bibinfo{year}{1972}).

\bibitem{Bender1996}
\bibinfo{author}{\bibfnamefont{J.}~\bibnamefont{Bender}},
  \bibnamefont{and} \bibinfo{author}{\bibfnamefont{N.~J.} \bibnamefont{Wagner}},
  \bibinfo{journal}{J. Chem. Phys.}
  \textbf{\bibinfo{volume}{40}}, \bibinfo{pages}{899}
  (\bibinfo{year}{1996}).

\bibitem{Maranzano2001}
\bibinfo{author}{\bibfnamefont{B.~J.}~\bibnamefont{Maranzano}},
  \bibnamefont{and} \bibinfo{author}{\bibfnamefont{N.~J.} \bibnamefont{Wagner}},
  \bibinfo{journal}{J. Rheol.}
  \textbf{\bibinfo{volume}{45}}, \bibinfo{pages}{1205}
  (\bibinfo{year}{2001}).
	
	\bibitem{Fall2008}
\bibinfo{author}{\bibfnamefont{A.}~\bibnamefont{Fall}},
  \bibinfo{author}{\bibfnamefont{N.}~\bibnamefont{Huang}},
  \bibinfo{author}{\bibfnamefont{F.}~\bibnamefont{Bertrand}},
  \bibinfo{author}{\bibfnamefont{G.}~\bibnamefont{Ovarlez}}, \bibnamefont{and}
  \bibinfo{author}{\bibfnamefont{D.}~\bibnamefont{Bonn}},
  \bibinfo{journal}{Phys. Rev. Lett.} \textbf{\bibinfo{volume}{100}},
  \bibinfo{pages}{018301} (\bibinfo{year}{2008}).
	
	\bibitem{Xu2014}
\bibinfo{author}{\bibfnamefont{Q.}~\bibnamefont{Xu}},
  \bibinfo{author}{\bibfnamefont{S.}~\bibnamefont{Majumdar}},
  \bibinfo{author}{\bibfnamefont{E.}~\bibnamefont{Brown}},
  \bibnamefont{and} \bibinfo{author}{\bibfnamefont{H.~M.}~\bibnamefont{Jaeger}},
  \bibinfo{journal}{ EPL (Europhysics Letters) } \textbf{\bibinfo{volume}{107}}, \bibinfo{pages}{68004}  (\bibinfo{year}{2014}).
		
	
	\bibitem{Guy2015}
\bibinfo{author}{\bibfnamefont{B.~M.}~\bibnamefont{Guy}},
  \bibinfo{author}{\bibfnamefont{M.}~\bibnamefont{Hermes}}
  \bibnamefont{and} \bibinfo{author}{\bibfnamefont{W.~C.~K.} \bibnamefont{Poon}},
  \bibinfo{journal}{Phy. Rev. Lett.}
  \textbf{\bibinfo{volume}{115}}, \bibinfo{pages}{088304}
  (\bibinfo{year}{2015}).
	
	
\bibitem{Royer2015}
\bibinfo{author}{\bibfnamefont{J.~R.}~\bibnamefont{Royer}},
  \bibinfo{author}{\bibfnamefont{D.~L.}~\bibnamefont{Blair}},
  \bibnamefont{and} \bibinfo{author}{\bibfnamefont{S.~D.} \bibnamefont{Hudson}},
  \bibinfo{journal}{Phys. Rev. Lett.}
  \textbf{\bibinfo{volume}{116}}, \bibinfo{pages}{188301}
  (\bibinfo{year}{2016}).
	
	\bibitem{Comtet2017}
\bibinfo{author}{\bibfnamefont{J.}~\bibnamefont{Comtet}},
  \bibinfo{author}{\bibfnamefont{G.}~\bibnamefont{Chatte}},
	\bibinfo{author}{\bibfnamefont{A.}~\bibnamefont{Nigues}},
	\bibinfo{author}{\bibfnamefont{L.}~\bibnamefont{Bocquet}},
	\bibinfo{author}{\bibfnamefont{A.}~\bibnamefont{Siria}},
  \bibnamefont{and} \bibinfo{author}{\bibfnamefont{A.} \bibnamefont{Colin}},
  \bibinfo{journal}{Nat. Commun.}
  \textbf{\bibinfo{volume}{8}}, \bibinfo{pages}{15633}
  (\bibinfo{year}{2017}).

\bibitem{Wagner2009}
\bibinfo{author}{\bibfnamefont{N.~J.}~\bibnamefont{Wagner}},
  \bibnamefont{and} \bibinfo{author}{\bibfnamefont{J.~F.}~\bibnamefont{Brady}},
  \bibinfo{journal}{Phys. Today}
  \textbf{\bibinfo{volume}{62}}, \bibinfo{pages}{27}
  (\bibinfo{year}{2009}).
	
	\bibitem{Singh2018}
\bibinfo{author}{\bibfnamefont{A.}~\bibnamefont{Singh}},
  \bibinfo{author}{\bibfnamefont{R.}~\bibnamefont{Mari}},
  \bibinfo{author}{\bibfnamefont{M.~M.} \bibnamefont{Denn}}
  \bibnamefont{and} \bibinfo{author}{\bibfnamefont{J.~F.} \bibnamefont{Morris}},
  \bibinfo{journal}{J. Rheol.}
  \textbf{\bibinfo{volume}{62}}, \bibinfo{pages}{405}
  (\bibinfo{year}{2018}).

\bibitem{Lee2003}
\bibinfo{author}{\bibfnamefont{Y.}~\bibnamefont{Lee}},
  \bibinfo{author}{\bibfnamefont{E.}~\bibnamefont{Wetzel}},
  \bibnamefont{and} \bibinfo{author}{\bibfnamefont{J.} \bibnamefont{Wagner}},
  \bibinfo{journal}{J. Mater. Sci.}
  \textbf{\bibinfo{volume}{38}}, \bibinfo{pages}{2825}
  (\bibinfo{year}{2003}).

\bibitem{Majumdar2013}
\bibinfo{author}{\bibfnamefont{A.}~\bibnamefont{Majumdar}},
  \bibinfo{author}{\bibfnamefont{B.}~\bibnamefont{Singh ~Butola}},
  \bibnamefont{and} \bibinfo{author}{\bibfnamefont{A.} \bibnamefont{Srivastava}},
  \bibinfo{journal}{Materials and Design}
  \textbf{\bibinfo{volume}{46}}, \bibinfo{pages}{191}
  (\bibinfo{year}{2013}).
		
		
	\bibitem{Lin2019}
\bibinfo{author}{\bibfnamefont{K.} \bibnamefont{Lin}},
  \bibinfo{author}{\bibfnamefont{H.}~\bibnamefont{Liu}},
	\bibinfo{author}{\bibfnamefont{M.}~\bibnamefont{Wei}},
	\bibinfo{author}{\bibfnamefont{A.}~\bibnamefont{Zhou}},
	  \bibnamefont{and} \bibinfo{author}{\bibfnamefont{F.} \bibnamefont{Bu}},       
  \bibinfo{journal}{Smart Mater. Struct.}
	\textbf{\bibinfo{volume}{28}}, \bibinfo{pages}{025007}
	(\bibinfo{year}{2019}).
	
	\bibitem{Qin2017}
\bibinfo{author}{\bibfnamefont{J.} \bibnamefont{Qin}},
  \bibinfo{author}{\bibfnamefont{G.}~\bibnamefont{Zhang}},
  \bibnamefont{and} \bibinfo{author}{\bibfnamefont{X.} \bibnamefont{Shi}},       
  \bibinfo{journal}{J. Dispers. Sci. Technol.} \textbf{\bibinfo{volume}{38}}, \bibinfo{pages}{935} (\bibinfo{year}{2017}).
	
		
	\bibitem{Chu2014}
\bibinfo{author}{\bibfnamefont{C. E.} \bibnamefont{Chu}},
  \bibinfo{author}{\bibfnamefont{J. A.}~\bibnamefont{Groman}},
	\bibinfo{author}{\bibfnamefont{H. L.}~\bibnamefont{Sieber}},
	\bibinfo{author}{\bibfnamefont{J. G.}~\bibnamefont{Miller}},
	\bibinfo{author}{\bibfnamefont{R. J.}~\bibnamefont{Okamoto}},
  \bibnamefont{and} \bibinfo{author}{\bibfnamefont{J. I.} \bibnamefont{Katz}},       
  \bibinfo{journal}{arXiv:1405.7233v1}
	(\bibinfo{year}{2014}).
	
	\bibitem{Majumdar2011}
\bibinfo{author}{\bibfnamefont{S.}~\bibnamefont{Majumdar}},
  \bibinfo{author}{\bibfnamefont{R.}~\bibnamefont{Krishnaswamy}},
  \bibnamefont{and} \bibinfo{author}{\bibfnamefont{A.~K.}~\bibnamefont{Sood}},
  \bibinfo{journal}{ Proc. Natl. Acad. Sci. U. S. A.} \textbf{\bibinfo{volume}{108(22)}}, \bibinfo{pages}{8996}  (\bibinfo{year}{2011}).

	
	\bibitem{Brown2014}
\bibinfo{author}{\bibfnamefont{E.}~\bibnamefont{Brown}} \bibnamefont{and}
  \bibinfo{author}{\bibfnamefont{H.~M.} \bibnamefont{Jaeger}},
  \bibinfo{journal}{Reports on Progress in Physics}
  \textbf{\bibinfo{volume}{77}}, \bibinfo{pages}{046602}
  (\bibinfo{year}{2014}).
	
	\bibitem{Fernandez2013}
\bibinfo{author}{\bibfnamefont{N.}~\bibnamefont{~Fernandez}},
  \bibinfo{author}{\bibfnamefont{~ R.}~\bibnamefont{~Mani}},
  \bibinfo{author}{\bibfnamefont{~ D.}~ \bibnamefont{~Rinaldi}},
	\bibinfo{author}{\bibfnamefont{~ D.}~ \bibnamefont{~Kadau}},
	\bibinfo{author}{\bibfnamefont{~M.} \bibnamefont{~Mosquet}},
	\bibinfo{author}{\bibfnamefont{~H.} \bibnamefont{~Lombois-Burger}},
	\bibinfo{author}{\bibfnamefont{~J.} \bibnamefont{~Cayer-Barrioz}},
	\bibinfo{author}{\bibfnamefont{~H.~J.} \bibnamefont{~Herrmann}},
	\bibinfo{author}{\bibfnamefont{~N.~D.} \bibnamefont{~Spencer}},
  \bibnamefont{and} \bibinfo{author}{\bibfnamefont{~L.} \bibnamefont{~Isa}},
  \bibinfo{journal}{Phys. Rev. Lett.}
  \textbf{\bibinfo{volume}{111}}, \bibinfo{pages}{108301}
  (\bibinfo{year}{2013}).
	
	\bibitem{Lin2015}
\bibinfo{author}{\bibfnamefont{N.~Y.~C.} \bibnamefont{Lin}},
  \bibinfo{author}{\bibfnamefont{B.~M.} \bibnamefont{Guy}},
  \bibinfo{author}{\bibfnamefont{M.}~\bibnamefont{Hermes}},
  \bibinfo{author}{\bibfnamefont{C.}~\bibnamefont{Ness}},
  \bibinfo{author}{\bibfnamefont{J.}~\bibnamefont{Sun}},
  \bibinfo{author}{\bibfnamefont{W.~C.} \bibnamefont{Poon}}, \bibnamefont{and}
  \bibinfo{author}{\bibfnamefont{I.}~\bibnamefont{Cohen}},
  \bibinfo{journal}{Phy. Rev. Lett.} \textbf{\bibinfo{volume}{115}},
  \bibinfo{pages}{228304} (\bibinfo{year}{2015}).
	
	
	\bibitem{Clavaud2017}
\bibinfo{author}{\bibfnamefont{C.}~\bibnamefont{Clavaud}},
  \bibinfo{author}{\bibfnamefont{A.}~\bibnamefont{Berut}},
	\bibinfo{author}{\bibfnamefont{B.}~\bibnamefont{Metzger}},
  \bibnamefont{and} \bibinfo{author}{\bibfnamefont{Y.} \bibnamefont{Forterre}},
  \bibinfo{journal}{Proc. Natl. Acad. Sci.(USA)}
  \textbf{\bibinfo{volume}{114}}, \bibinfo{pages}{5147}
  (\bibinfo{year}{2017}).
	
	
\bibitem{Seto2013}
\bibinfo{author}{\bibfnamefont{R.}~\bibnamefont{Seto}},
  \bibinfo{author}{\bibfnamefont{R.}~\bibnamefont{Mari}},
  \bibinfo{author}{\bibfnamefont{J.~F.} \bibnamefont{Morris}},
  \bibnamefont{and} \bibinfo{author}{\bibfnamefont{M.~M.} \bibnamefont{Denn}},
  \bibinfo{journal}{Phys. Rev. Lett.} \textbf{\bibinfo{volume}{111}},
  \bibinfo{pages}{218301} (\bibinfo{year}{2013}).

\bibitem{Mari2015}
\bibinfo{author}{\bibfnamefont{R.}~\bibnamefont{Mari}},
  \bibinfo{author}{\bibfnamefont{R.}~\bibnamefont{Seto}},
  \bibinfo{author}{\bibfnamefont{J.~F.} \bibnamefont{Morris}},
  \bibnamefont{and} \bibinfo{author}{\bibfnamefont{M.~M.} \bibnamefont{Denn}},
  \bibinfo{journal}{J. Rheol}
  \textbf{\bibinfo{volume}{58}}, \bibinfo{pages}{1693}
  (\bibinfo{year}{2014}).
	
	\bibitem{Smith2010}
\bibinfo{author}{\bibfnamefont{M.}~\bibnamefont{Smith}},
  \bibinfo{author}{\bibfnamefont{R.}~\bibnamefont{Besseling}},
  \bibinfo{author}{\bibfnamefont{M.}~\bibnamefont{Cates}}, \bibnamefont{and}
  \bibinfo{author}{\bibfnamefont{V.}~\bibnamefont{Bertola}},
  \bibinfo{journal}{Nature Communications} \textbf{\bibinfo{volume}{1}},
  \bibinfo{pages}{114} (\bibinfo{year}{2010}).
	
	
\bibitem{Peters2016}
\bibinfo{author}{\bibfnamefont{I.~R.}~\bibnamefont{Peters}},
  \bibinfo{author}{\bibfnamefont{S.}~\bibnamefont{Majumdar}}
  \bibnamefont{and} \bibinfo{author}{\bibfnamefont{H.~M.} \bibnamefont{~Jaeger}},
  \bibinfo{journal}{Nature}
  \textbf{\bibinfo{volume}{532}}, \bibinfo{pages}{214}
  (\bibinfo{year}{2016}).
	
	\bibitem{Majumdar2017}
\bibinfo{author}{\bibfnamefont{S.}~\bibnamefont{Majumdar}},
  \bibinfo{author}{\bibfnamefont{I.~R.}~\bibnamefont{Peters}},
  \bibinfo{author}{\bibfnamefont{E.}~\bibnamefont{Han}},
  \bibnamefont{and} \bibinfo{author}{\bibfnamefont{H.~M.}~\bibnamefont{Jaeger}},
  \bibinfo{journal}{Phys. Rev. E } \textbf{\bibinfo{volume}{95}}, \bibinfo{pages}{012603}
  (\bibinfo{year}{2017}).

	
\bibitem{James2018}
\bibinfo{author}{\bibfnamefont{N.~M.}~\bibnamefont{James}},
  \bibinfo{author}{\bibfnamefont{E.}~\bibnamefont{Han}},
  \bibinfo{author}{\bibfnamefont{R.~A.} \bibnamefont{Lopez De La Cruz}},
	\bibinfo{author}{\bibfnamefont{J.}~\bibnamefont{Jureller}}
  \bibnamefont{and} \bibinfo{author}{\bibfnamefont{H.~M.}~\bibnamefont{Jaeger}},
  \bibinfo{journal}{Nat. Mat.}
  \textbf{\bibinfo{volume}{17}}, \bibinfo{pages}{965}
  (\bibinfo{year}{2018}).
	
\bibitem{Han2018}
\bibinfo{author}{\bibfnamefont{E.}~\bibnamefont{Han}},
  \bibinfo{author}{\bibfnamefont{N.~M.}~\bibnamefont{James}}
  \bibnamefont{and} \bibinfo{author}{\bibfnamefont{H.~M.} \bibnamefont{Jaeger}},
  \bibinfo{journal}{arXiv:1810.11887v1}
	(\bibinfo{year}{2018}).
  
	
\bibitem{Wyart2014}
\bibinfo{author}{\bibfnamefont{M.}~\bibnamefont{Wyart}},
   \bibnamefont{and} \bibinfo{author}{\bibfnamefont{M.~E.}~\bibnamefont{Cates}},
  \bibinfo{journal}{Phy. Rev. Lett.} \textbf{\bibinfo{volume}{114}},
  \bibinfo{pages}{098302} (\bibinfo{year}{2014}).
	

\bibitem{Krieger1957}
\bibinfo{author}{\bibfnamefont{I.~M.}~\bibnamefont{Krieger}},
  \bibnamefont{and} \bibinfo{author}{\bibfnamefont{T.~J.}~\bibnamefont{Dougherty}},
  \bibinfo{journal}{Trans. Soc. Rheol.}
  \textbf{\bibinfo{volume}{3}}, \bibinfo{pages}{137}
  (\bibinfo{year}{1957}).
	


	\bibitem{Nicolas2018}
\bibinfo{author}{\bibfnamefont{A.} \bibnamefont{Nicolas}},
  \bibinfo{author}{\bibfnamefont{E. E.}~\bibnamefont{Fererro}},
	\bibinfo{author}{\bibfnamefont{K.}~\bibnamefont{Martens}},
	  \bibnamefont{and} \bibinfo{author}{\bibfnamefont{J-L.} \bibnamefont{Barrat}},       
  \bibinfo{journal}{Rev. Mod. Phys.}
	\textbf{\bibinfo{volume}{90}}, \bibinfo{pages}{045006}
  (\bibinfo{year}{2018}).
	
	
	
	\bibitem{Howell1999}
\bibinfo{author}{\bibfnamefont{D.}~\bibnamefont{Howell}},
  \bibinfo{author}{\bibfnamefont{R.~P.}~\bibnamefont{Behringer}},
	\bibnamefont{and}\bibinfo{author}{\bibfnamefont{C.}~\bibnamefont{Veje}},
    \bibinfo{journal}{Phys. Rev. Lett.}
  \textbf{\bibinfo{volume}{82}}, \bibinfo{pages}{5241}
  (\bibinfo{year}{1999}).
	
	
	\bibitem{O'Harn2001}
\bibinfo{author}{\bibfnamefont{C.~S.}~\bibnamefont{O'Harn}},
  \bibinfo{author}{\bibfnamefont{S.~A.}~\bibnamefont{Langer}},
	\bibinfo{author}{\bibfnamefont{A.~J.}~\bibnamefont{Liu}},
  \bibnamefont{and} \bibinfo{author}{\bibfnamefont{S.~R.} \bibnamefont{Nagel}},
  \bibinfo{journal}{Phys. Rev. Lett.}
  \textbf{\bibinfo{volume}{86}}, \bibinfo{pages}{111}
  (\bibinfo{year}{2001}).

	
	\bibitem{Guy2019}
\bibinfo{author}{\bibfnamefont{M.~M}~\bibnamefont{Guy}},
  \bibinfo{author}{\bibfnamefont{C.}~\bibnamefont{Ness}},
	\bibinfo{author}{\bibfnamefont{M.}~\bibnamefont{Hermes}},
  \bibinfo{author}{\bibfnamefont{L.~J.}~\bibnamefont{Sawiak}},
	\bibinfo{author}{\bibfnamefont{J.}~\bibnamefont{Sun}},
  \bibinfo{author}{\bibfnamefont{W.~C.} \bibnamefont{Poon}},
  	\bibinfo{journal}{arXiv:1901.02066v1}
	(\bibinfo{year}{2019}).
	
	
	\bibitem{Silbert2010}
\bibinfo{author}{\bibfnamefont{L.~E.}~\bibnamefont{Silbert}},
    \bibinfo{journal}{Soft Matter}
  \textbf{\bibinfo{volume}{6}}, \bibinfo{pages}{2918}
  (\bibinfo{year}{2010}).
	
	\bibitem{AntonPaar}
	\bibinfo{website}{https://wiki.anton-paar.com/en/the-influence-of-particles-on-suspension-rheology/}
	
	\bibitem{Maranzano2001_1}
\bibinfo{author}{\bibfnamefont{B.~J.}~\bibnamefont{Maranzano}},
  \bibnamefont{and} \bibinfo{author}{\bibfnamefont{N.~J.} \bibnamefont{Wagner}},
  \bibinfo{journal}{J. Chem. Phys.}
  \textbf{\bibinfo{volume}{114}}, \bibinfo{pages}{10514}
  (\bibinfo{year}{2001}).
	
	
	
	\bibitem{Brown2010}
\bibinfo{author}{\bibfnamefont{E.}~\bibnamefont{Brown}},
  \bibinfo{author}{\bibfnamefont{N.~A.}~\bibnamefont{Forman}},
  \bibinfo{author}{\bibfnamefont{C.~S.} \bibnamefont{Orellana}},
	\bibinfo{author}{\bibfnamefont{H.}~\bibnamefont{Zhang}}
	\bibinfo{author}{\bibfnamefont{B.~W.}~\bibnamefont{Maynor}}
	\bibinfo{author}{\bibfnamefont{D.~E}~\bibnamefont{Betts}}
	\bibinfo{author}{\bibfnamefont{J.~M.}~\bibnamefont{DeSimone}}
  \bibnamefont{and} \bibinfo{author}{\bibfnamefont{H.~M.}~\bibnamefont{Jaeger}},
  \bibinfo{journal}{Nat. Mat.}
  \textbf{\bibinfo{volume}{9}}, \bibinfo{pages}{220}
  (\bibinfo{year}{2010}).

\end{thebibliography}

\vspace{1 cm}
\begin{thebibliography}{24}
\expandafter\ifx\csname natexlab\endcsname\relax\def\natexlab#1{#1}\fi
\expandafter\ifx\csname bibnamefont\endcsname\relax
  \def\bibnamefont#1{#1}\fi
\expandafter\ifx\csname bibfnamefont\endcsname\relax
  \def\bibfnamefont#1{#1}\fi
\expandafter\ifx\csname citenamefont\endcsname\relax
  \def\citenamefont#1{#1}\fi
\expandafter\ifx\csname url\endcsname\relax
  \def\url#1{\texttt{#1}}\fi
\expandafter\ifx\csname urlprefix\endcsname\relax\def\urlprefix{URL }\fi
\providecommand{\bibinfo}[2]{#2}
\providecommand{\eprint}[2][]{\url{#2}}

	
\bibitem{Wang2013}
\bibinfo{author}{\bibfnamefont{D.} \bibnamefont{Wang}},
  \bibinfo{author}{\bibfnamefont{B.}~\bibnamefont{Yu}},
  \bibinfo{author}{\bibfnamefont{H.~L.} \bibnamefont{Cong}},
  \bibinfo{author}{\bibfnamefont{Y.~Z.} \bibnamefont{Wang}},
  \bibinfo{author}{\bibfnamefont{Q.} \bibnamefont{Wu}},
  \bibnamefont{and} \bibinfo{author}{\bibfnamefont{J.~L.} \bibnamefont{Wang}}, 
  \bibinfo{journal}{ Integr Ferroelectr} \textbf{\bibinfo{volume}{147(1)}}, \bibinfo{pages}{41}    (\bibinfo{year}{2013}).

\bibitem{Paine1990}
\bibinfo{author}{\bibfnamefont{A.~J.} \bibnamefont{Paine}},
  \bibinfo{author}{\bibfnamefont{W.}~\bibnamefont{Luymes}},
  \bibnamefont{and} \bibinfo{author}{\bibfnamefont{J.} \bibnamefont{Mcnulty}},       
  \bibinfo{journal}{Integr Ferroelectr} \textbf{\bibinfo{volume}{23}}, \bibinfo{pages}{3104} (\bibinfo{year}{1990}).


\bibitem{Cho2016}
\bibinfo{author}{\bibfnamefont{Y.~S.} \bibnamefont{Cho}},
  \bibinfo{author}{\bibfnamefont{C.~H.}~\bibnamefont{Shin}},
  \bibnamefont{and} \bibinfo{author}{\bibfnamefont{S.} \bibnamefont{Han}},       
  \bibinfo{journal}{ Nanoscale Res. Lett.} \textbf{\bibinfo{volume}{11}}, \bibinfo{pages}{46} (\bibinfo{year}{2016}).


\end{thebibliography}
\end{document}